# Using Design Science to Build a Watermark System for Cloud Rightful Ownership Protection

**Brian Cusack**
School Mathematics & Computer Science
AUT University
Auckland
Email: brian.cusack@aut.ac.nz

**Reza Khaleghparast**
School Mathematics & Computer Science
AUT University
Auckland
Email: rkhalegh@aut.ac.nz

## Abstract

Cloud computing opportunities have presented service options for users that are both economical and flexible to use requirements. However, the risk analysis for the user identifies vulnerabilities for intellectual property ownership and vulnerabilities for the identification of rightful property owners when cloud services are used. It is common for image owners to embed watermarks and other security mechanisms into their property so that the rightful ownership may be identified. In this paper we present a design that overcomes many of the current limitations in cloud watermarking uses; and propose a schema that places responsibility on the cloud provider to have a robust information protection program. Such a design solution lays out an information security architecture that enhances utility for cloud services and gives better options for users to securely place properties in the cloud. The Design Science methodology is used to build the artefact and answer the research question: How can rightful ownership be protected in the Cloud?

## 1 Introduction

One of the important issues of cloud computing is user loss of control. The system architecture for services posits multiple layers of inter-related services for which no one supplier has control. Figure 1 shows the technical services stack (Tek, et al., 2010, p.684) and figure 2 the service architecture referred in this problem statement (Tek, et al., 2010, p.686). In the first instance a user interacts with a sales agent (human or machine) to purchase the services opportunity. The sales agent may be selling on behalf of one or more service suppliers. In turn these suppliers have supply agreements with many sub-service suppliers or brokers. Sub-service suppliers also have inter-related arrangements for services that may migrate data and service without notice (Lombardi and Di Pietro, 2011). The net result is that a cloud service user may not know the storage and processing place or places of the data and may not be assured of ownership protection. Hence, the consequences are for security, privacy and legal jurisdiction. The essence of cloud computing is that a user entrusts their own digital information to a second party who exploits multiple third parties to deliver the user a service.

| **CRM & ERP** | **Software as a Service** | **Email, Office, User Interface** |
|---|---|---|
| | **Platform as a Service** | Design, Modelling, Development, Testing |
| Computing, Storage, Communications | **Infrastructure as a Service** | |
| | **Data Centres** | |
| | **Everything as a Service** | |

*Figure 1: Cloud Computing Services (Based on: Tek, et al., 2010; Mel and Grace, 2011)*

The user has technology and information, which are hosted in the cloud by the provider, and the services to store information, to create further information, and to transact business are made available by the provider. Inevitably, the protection of ownership rights is an issue and the many related vulnerabilities require risk treatment in a secure service system (O'Ruanaidh, 1996; Cayre, 2005).

In this paper the problem is addressed by reviewing the potential of watermarks to protect rightful ownership and to place the responsibility for that protection with the service provider. Another example of the service users losing control is the scope of service level agreements (SLAs) and the enforceability between cloud providers (Lombardi and Di Pietro, 2011). Security of Cloud computing has been enhanced in many ways, and improved with for example, watermarking. Watermarking is a technology for copyright protection that mitigates illegally copying or tampered. It introduces small patterns in the digital data signal without changing the original source. If there is a breach of the original data then, the rightful owner can verify the ownership of that data (Liu et al., 2011). It is used to protect visibly or invisibly the ownership of artefacts such as images, audios and, videos. Currently there are many software packages available for users to insert digital watermarks in their media.

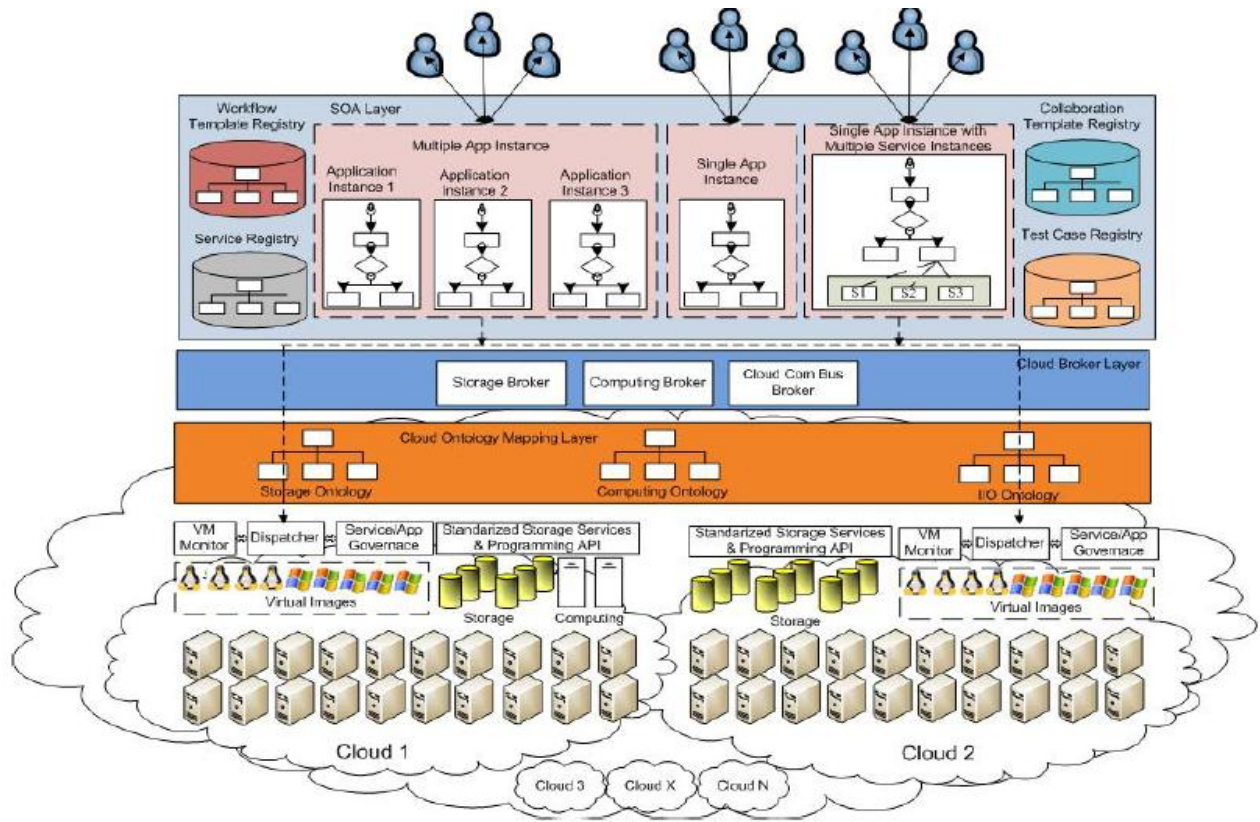

*Figure 2: Cloud Service Architecture (*Tek, et al., 2010, p.686)

It is the concern in this paper that the user insertion of watermarks may have variable impact on the problem of verifying rightful ownership (Sherekar, 2008). The cloud environment is a torrid environment in which there are many possible attacks that may be on account of unintentional management of the data or intentional attacks on the data. The variation introduced by many different user watermarking tools can be reduced by requiring cloud service providers to insert watermarks.

The problem partial solution transfers responsibility to the service provider to have a robust and consistent capability for watermarking. Consequently, a secure information management service by a provider is required to test and prove watermarks robust to the environment in which the service is provided. Users applying generic watermark tools may not have the capability to anticipate the scope of attacks a property may be subjected and the management practices of multiple third parties. Hence we advocate an architecture where the responsibility is with the service provider and show a tool design for provider information security management. This paper is structured to review previous literature on cloud architecture and watermarking, provide a methodology for research (Henver et al., 2004), to demonstrate a design solution and to evaluate the solution.

## 2   Previous Literature

A select review of relevant literature relating to defining Cloud Computing, the related security issues and Watermarking is made. The brief review provides a context for the problem and the proposed solution.

### 2.1   Cloud Definitions

Cloud computing has been defined in various ways for instance, Furht (2010, p.3) defined Cloud computing as, "a new style of computing in which dynamically scalable and often virtualized resources are provided as a services over the Internet". According to Mollah, et al. (2012, p1), Cloud computing is a, "TCP/IP based high development and integrations of computer technologies such as fast microprocessor, huge memory, high-speed network and reliable system architecture." The National Institute of Standards and Technology in Special Publication 800-145 defines Cloud computing as "a model for enabling ubiquitous, convenient, on-demand network access to a shared pool of configurable computing resources (e.g., networks, servers, storage, applications, and services) that can be rapidly provisioned and released with minimal management effort or service provider interaction" (Mell and Grance, 2011, p.3). The cloud is said to be a network of data-centers working together to provide powerful applications, platforms and services that can be accessed by its users over the Internet. Figure 1 shows the layering of technical services that constitute a cloud environment.

Cloud computing has four main deployment models. The first model is known as "Private Cloud". This model refers to a cloud infrastructure that may be owned and operated by an organization for private use only. The second model is known as "Community Cloud". This model refers to a cloud infrastructure that is owned, managed and used exclusively by a community with similar concerns such as security requirements, policies or mission (Wei-Tek, et al., 2010). The third model is called "Public Cloud", this model refers to an infrastructure that is open to the general public. The infrastructure may be owned and operated by an academic institution, government organization or a business. The final cloud model is called "Hybrid Cloud", this model refers to an infrastructure which is a combination of two or more of the other three models. The particular model allows the infrastructure to remain exclusive while bound by standards or branded technologies. Figure 2 illustrates the complexity of the layered cloud architectures and the multiple choices a user may make in choosing service. The users are depicted at the top of the figure and the layers beneath provide the pay-as-you-go services. Intermediation by infra-structure is provided by the mapping of the user requirement to the virtualized domains of service. Importantly at the bottom of the figure the inter-related nature of cloud service providers is illustrated indicating the complexity of the relationship of data and services. The user is separated from their data in many potential ways and the ownership of that data transfers to multiple providers who provide processing services and storage (Cayre et al., 2005).

### 2.2   Cloud Security

The expected benefits of Cloud Computing provide user motivation for risk taking. The services have a high potential for cost reduction, improving productivity, agility, flexibility and greater convenience to the users. However, the benefits require weighting against the many unresolved issues for data security, provider verification, privacy protection, regulation and the jurisdictional barriers (Ruan et al., 2013). The unresolved issues raise serious threats and challenges for cloud computing and its users (Chuhong, et al., 2006). The essence of cloud computing is that a user entrusts the digital information to the cloud computing service provider in the belief the other related parties will protect the owner interest. All these steps happen online with technology which the user has little or no control over. Inevitably, ownership rights are affected (Liu et al., 2011). Multimedia works such as images, audios and, videos have copyright and integrity concerns, and identification requirements once they are released into the Cloud.

To ensure only authorized users logon to cloud applications, multi-factor authentication is required. This is more than authentication based on what you know (username and password) and a second factor based on what you have: a one-time password, is required. The multi-factor strong authentication can be provided in a number of ways. First, a Hardware Token, which uses a dedicated device, such as an RSA SecureID token. Second, Software Token, which based on something the user (employee, contractor, customer, and business partner) probably already have. Third, uses physical characteristic (iris, fingerprint, and voiceprint) of the user. Previous work on image authentication falls into two groups, digital signatures and digital watermarks (Cayre et al., 2005). A digital signature is based upon the idea of public key encryption. A private key is used to encrypt a hashed version of the image. This encrypted file then forms a unique "signature" for the image since only the entity signing the image has knowledge of the private key used. An associated public key can be used to decrypt the signature. The image in question can be hashed using the same hashing function as used originally. If these hashes match then the image is authenticated. Digital signatures can be used for more than just image authentication. In particular when combined with secure timestamp, a digital signature can be used as a proof of first authorship. A watermark, on the other hand, is a code secretly embedded into the image. The watermark allows for verification of the origin of an image. However, a watermark alone is not enough to prove first authorship, since an image could be marked with multiple watermarks (Chuhong et al., 2006).

The main security concerns of Cloud computing relate to the planned management of information and the unplanned unauthorised access and use of information. Both constitute attacks and present vulnerabilities. In the first instance the lack of standards, standardisation and interoperability agreements leads to the spoliation of information by the way it is processed by multiple service providers. For example some providers crop or resize images, others compress text and so on as part of the information management policies (For example, https://www.facebook.com/help/266520536764594). These actions can change the information in ways that compromise the integrity of the information and the ability of the owner to verify ownership when for example hashes are changed, watermarks destroyed and so on. In the second instance the Cloud is vulnerable to many of the well-known networking attacks such as flooding, spoofing, hijacking, and so on. However the architectural design and processes of the Cloud also adds other potential attacks such as injection, rollbacks, wrapping, cache diving, side-channel and so on (Sherekar et al., 2011). As a consequence information protection requires all the traditional mechanisms of network security but also new mechanisms for the Cloud environment (Lui et al., 2011).

## 2.3 Watermarking

Digital watermarking is a solution for rightful ownership identification when data and owners are separated by system. Watermarking can be implemented to make a safer way for data transfer protection (Yang et al., 2011; Yu et al., 2011). A major problem faced by content providers and owners is protection of their material. They are concerned about copyright protection and other forms of abuse of their digital content. Unlike copies of analogue tapes, copies of digital data are identical to the original and suffer no quality degradation, and there is no limit to the number of exact copies that can be made. In addition, digital equipment that can make digital copies is widely available and inexpensive. One approach to content security uses cryptographic techniques, but those encryption systems do not completely solve the problem of unauthorized copying (Yang et al., 2011). All encrypted content needs to be decrypted before it can be used. Once encryption is removed, there is no way to prove the ownership or copyright of the content. As a solution to this problem digital watermark technology provides protection against unauthorized copying of digital content. A digital watermark is a signal added to the original digital data (namely, audio, video, or image), which can later be extracted or detected. The watermark has intended to be permanently embedded into the digital data so that authorized users can easily access it. At the same time, the watermark should not degrade the quality of the digital data. In general, digital watermark techniques must satisfy the following requirements (Won and Woo, 2001).

A digital watermark can be either visible or invisible. An example of digital visible watermark is the translucent logos that are often seen embedded in the corner of videos or images, in an attempt to prevent copyright infringement. However, these visible watermarks can be targeted and removed rather simply by cropping the media, or overwriting the logos. Subsequently, the field of digital watermarking is primarily focused on embedding invisible watermarks, which operate by tweaking the content of the media imperceptibly. As the watermark cannot be seen, there must exist a robustness property that ensures the watermark data survives if the image is altered (Johnson et al., 2001). Typical applications of digital watermarking can include broadcast monitoring, owner identification, proof of ownership, transaction tracking, content authentication, copy control, device control legacy enhancement and content description. The watermarked work is produced from an embedding algorithm that is traditionally comprised of three inputs: the original work, the watermark and a key. The watermark is extracted from the watermarked work in the blind process by using a detection algorithm in conjunction with the same key that was originally used to embed the watermark. In contrast, in a non-blind (or informed) watermark detection process that extracts the watermark the original work has to be provided as a reference source in order for the detection algorithm to function (Zhu and Hu, 2008).

## 3   Research Methodology

Design Science (DS) is an organising framework and philosophy for making and building artefacts. It has been made relevant to Information Systems (IS) research as a methodology and in this paper we apply the framework to IS security (Hevner, et al., 2004; Nunamaker, et al., 1990; Goes, 2014). The benefit of the approach is that an artefact may be investigated in context and the artefact improved through continuous iterations and testing (Walls, et al., 2004). The purpose of the DS research methodology is not only to develop an artefact but also to answer research questions. Depending on the characteristics and the goals of the research, a researcher can shape the processes to deliver innovative or confirmatory outcomes (Johannesson and Perjons, 2014). The DS research methodology consists of six main phases: problem identification and motivation, define the objectives for a solution, design and development, demonstration, evaluation and communication as it is shown in figure 3.

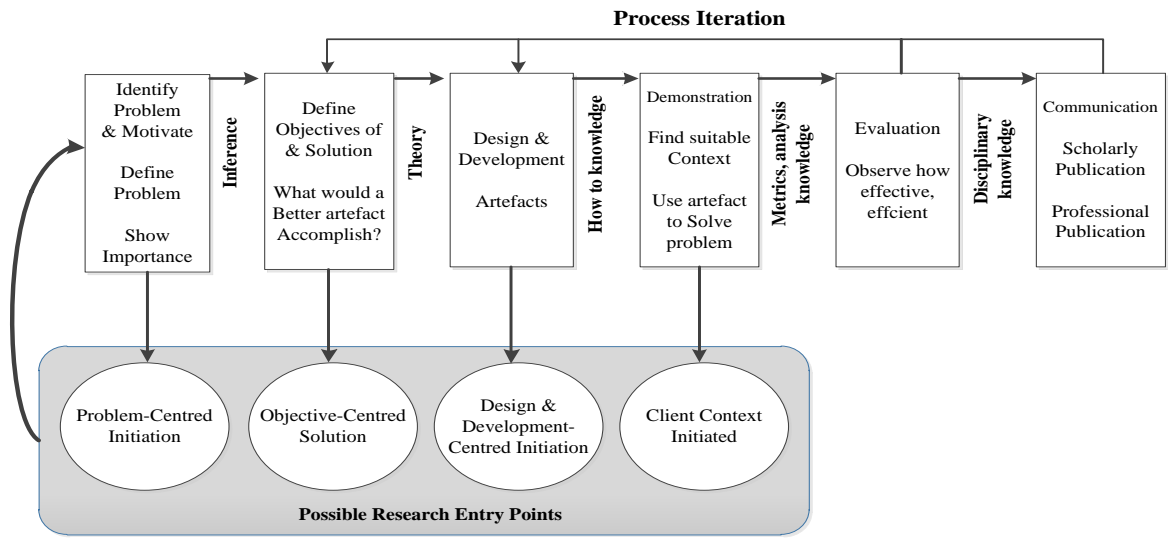

*Figure 3: DS research methodology (Peffers, et al., 2007, p.54)*

DS is solution oriented whereas the other research methodology such Natural Science or Social Science, are problem oriented (Hevner and Chatterjee, 2010). Figure 3 shows four entry points for starting research and six phases that are linked by output loops and feedback loops. The consequence is that any action that is taken is balanced by evaluation and the outcome of the evaluation can deliver forward propagation to the next phase or a return to an earlier phase for improvement. The first four phases also offer the option of returning to the entry specification for improvement and then re-entry to the phases. Phases 5 and 6 have process iteration options for quality improvement that offer alternative pathways depending on the researcher objectives and intended delivery standard.

In this research the six phases are defined as:

- Identify the Problem
- Define the Solution
- Design and Develop the Artefact
- Demonstrate in Context
- Evaluate the Solution
- Communicate the Story

Design Science is chosen for this study because it is solution oriented and not problem oriented. The problem specification in the Introduction and the literature analysed shows that the problem has two components. One technical and one managerial. DS focuses on the creation process and refining of the artefact to get a working solution. The purpose of this study is to develop a solution for assuring the rightful ownership of a property in a cloud environment. According to Offermann, et al., (2009, p.2), design science refers to "an explicitly organised, rational and wholly systematic approach to design; not just the utilisation of scientific knowledge of artefacts". Therefore, the solution defined is in two parts; one that addresses a requirement for information security and the other for an information security management design.

The design and development of the artefact concerns the technical solution for a robust watermark. The scope of the current research is to subject the solution to five attacks that represent information management policies in the Cloud. The two components of the solution are dependant whereby the managerial design solution solves the problem of user variation and the problem of watermark failure on account of user capability. The technical watermark artefact development is a proposed solution to technical failure. It has a reasoned layering of protection from information management attacks and a scope (that is untested in this research) for Cloud technical attacks. A server side rightful detection tool requires that every file coming to the server is assessed for consistency with the criteria for a robust watermark in the cloud environment. Any incoming file not meeting the requirement is then deleted and replaced by a service provider one. In this proposed research a context and a scope is selected that is feasible for testing. The scope of watermark research is narrowed to image media; JPG format; invisible perceptivity; robust requirements; image type; frequency domain processing; DWT format; and, private information for extraction. To satisfy the scope ten files were subjected to attack. The ten images were chosen as the cover objects for watermarks and were publicly available for free download. The scoping of the testing allowed the information management attacks of resizing, cropping, format change, text manipulation and flipping. Each of these attacks was chosen to represent standard policies applied by Cloud providers rather than for any complex malicious attacks that may exist in the cloud (these are out of scope). Once attacked and entered into the cloud database the images were extracted and tested for responsiveness to the original key and consistency against the original watermark. The PSNR scale was used for measuring the extracted watermark signal strength and the benchmark of less than 30 decibels selected from literature as a spoiled watermark (Oligeri et al., 2011).

The scope of the testing is to demonstrate the artefact in action in a simulated Cloud environment and in the context of information management attack. The simulation consisted of the artefact, the service provider policies, the information management attacks, a Cloud database, the embedding and extraction algorithms, and a PSNR measurement tool. As a consequence the demonstration provides a confirmation of the expectations an intellectual property owner may have for rightful ownership protection in similar circumstances. The evaluation is guided by the scope of the testing outlined here and cannot be generalised to matters outside of this scope. The final phase defined is the communication of the research findings and story. The phase is completed in the reporting of the results below and any other publications that may arise (Gregor and Hevner, 2013).

## 4   The Results

The testing proceeded in accordance with the limitations and constraints outlined above. The following sub-sections report the outcomes in each of the DS phases completed.

### 4.1   Identify the Problem

The first entry point is known as the problem-centred initiation (see figure 3). This entry point is designed to identify the problem and the motivation. In this case the identification came from a literature search of the topics of Cloud, Cloud security and watermarking. The problem was resolved into two elements; one that concerned the secure management of information and the other that concerned the robust build of a watermark that would remain resilient under five attacks. The relevance and importance of the study had been established in the Cloud literature on information security. The expected outcome is an addition to the current state of knowledge and the confirming or opening of starting points for further research.

### 4.2   Define a Solution

The second entry point is known as the objective-centred solution (see figure 3). This entry point is designed to support the designing of the artefact and the supporting literature research. The solution was determined in two dependant dimensions; one for information management security and the other for IT security. The resolution consequently impacted the problem as a comprehensive but partial solution. This was a deliberate ploy to make the testing achievable and proof of concept feasible. This phase was adequately documented in the literature review and scoped in the methodology section so that the defined solution acted as a target or a goal to achieve in the research.

### 4.3   Design and Develop Artefact

Design and development-centred initiated entry point or phase three of the DS research methodology is concerned with the creation of the artefact. The watermark artefact had three principle components; the embedding algorithm, the extraction algorithm and the three security features. The preparation algorithm was integrated with the development of the security features and the management context so that once the security features were stable then the embedding algorithm could add these features to the image as a watermark. Together the three features formed the basis of the artefact. The preparation algorithm also provided the link between the technical and management components of the solution. In the first decisions of the flow diagram (figure 4) a determination of the status of the incoming image is made to address the issue of user watermarks verses service provider watermarks. In figure 5 the logical steps for embedding the watermark are given. The embedding process must consider the three channels of red, blue and green that form the basis of image colour. By frequency blue is chosen first (a lower frequency signal) to enact the embedding process pixel by pixel. Red and green then follow to pick up the extra payload of a watermark. In figure 6 the extraction process is described.

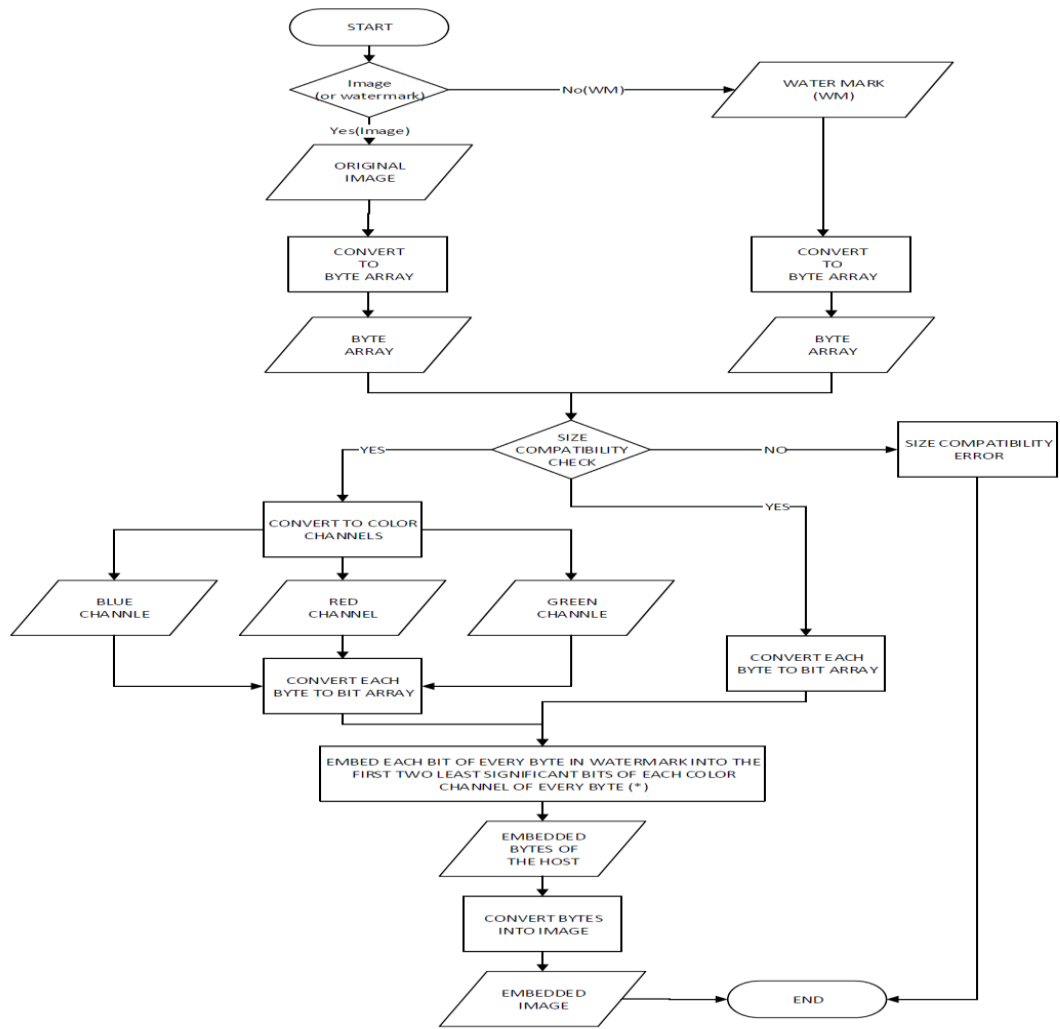

*Figure 4: Preparation Flow Chart*

Here the watermark signal must be detected and then tested for damage. The extracted watermark is evaluated against the input watermark for the purpose of testing. In the real world the evaluation would simply be against signal strength for tampering detection and against the security features for authentication. In this way the rightful ownership may be determined and with reference to a signature database.

The three security features that form the core to the artefact were constructed from data available in the cloud environment to uniquely identify the user. The first feature termed ISCH allows an image a user uploads to be stored with original hex and hash tags. The second feature termed CFDH comprises of a fixed password, a dynamic password and a hash. The CFDH consequently provides unique identification that is carried in the watermark. The third feature is the watermark existence check that is outlined in figure 6. Together these security features provide unique identification for the user in the uploading action, in the Cloud processing and in the Cloud database.

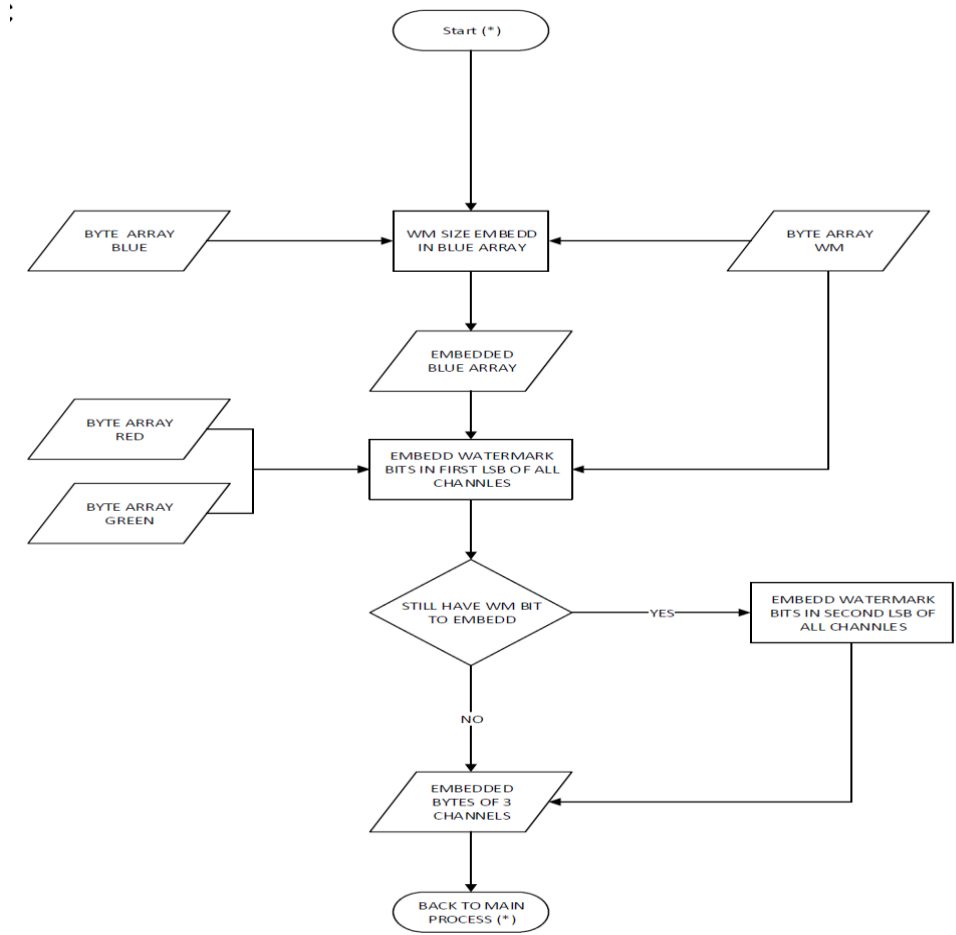

*Figure 5: Embedding algorithm*

## 4.4 Demonstration

The fourth phase of the DS research methodology requires a demonstration of the artefact in context and an assessment of the ability to solve the problem. This was achieved by preparing the artefact with the three unique security features. The features were then embedded into the ten test images as invisible watermarks. Each image was passed through the embedding algorithm (figure 5) and then stored in a Cloud database. The Cloud database was hosted on eight processors each with eight cloud environments. The images were transacted between environments and attacked by the five management attacks sequentially and in combinations until a sample data set of images were available for analysis. For analysis the images were extracted using the extraction algorithm (figure 6) and the means and standard deviations calculated for each image. The results showed that the artefact is a solution to the problem of rightful ownership and to a large extent the data shows control can be retained by the user when an intellectual property is entrusted to multiple service suppliers.

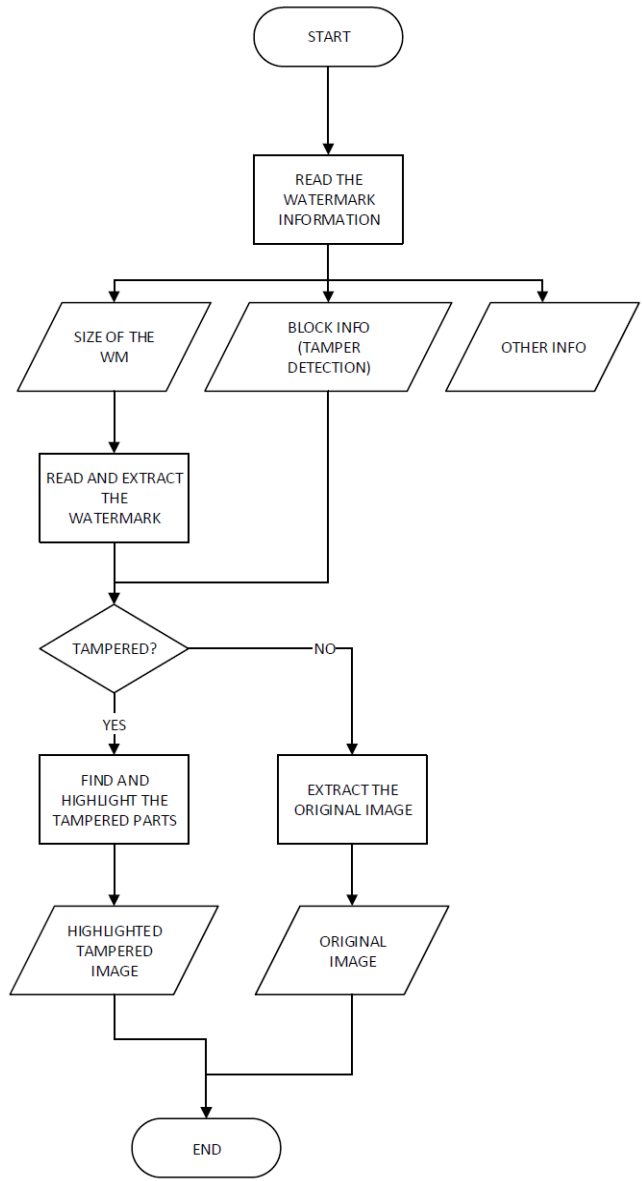

*Figure 6: Extraction algorithm*

## 4.5 Evaluation

Evaluation is the fifth phase of the DS research methodology; this involves observing and evaluating how effective and how efficient that the artefact solves a problem. In this phase, the evaluation and observation results from the entry point number four will be compared with the objectives of a solution. The results show that the artefact performed well under testing and the proposal is a viable solution. At the end of this phase, the researcher can then decide whether to iterate back to the entry point number 3 to improve the effectiveness of the artefact. In this instance the watermarks performed above expectation and to an acceptable level for use but what is not tested in this research is the broader range of Cloud technical attacks. Iteration to entry point 3 would redress the concern and bring quality improvement.

To evaluate the resilience of the standard watermark against attacks the PSNR signal to noise ratio was used for each image. It was calculated for each image for each attack when the image was retrieved from the cloud database. The PSNR ratio is usually set to 30 decibels (dB) or above as acceptance the watermark is verifiable. Clearly the higher the PSNR number the better the quality of the watermark and the clarity of its features. The following Table 1 illustrates the PSNR of the ten test images with embedded watermarks, which have been passed through the cloud environment and used in this research.

| NO. | IMAGE NAME | PSNR OF W.IMAGE |
|---|---|---|
| 1 | Cameraman | 71 |
| 2 | House | 74 |
| 3 | Jet plane | 64 |
| 4 | Lake | 57 |
| 5 | Lena | 64 |
| 6 | Living Room | 59 |
| 7 | Mandrill | 65 |
| 8 | Peppers | 59 |
| 9 | Pirate | 60 |
| 10 | Bridge | 52 |

*Table 1: Watermark Performance*

The results show that some loss is incurred in the attacks simulating policy impacts in a cloud environment. The PSNR values all show more than 50, which is greater than the 30 dB cut off for rejecting an unverifiable watermark. The greatest variance is between images that have the widest distribution of pixels suggesting that uniform images with relatively dense pixels provide the best cover objects.

### 4.6   Communication

The final phase of DS research methodology is known as Communication. This phase is designed to allow the researcher to employ various scholarly outlets to communicate the outcome of the study. This publication communicates the six phases completed and the finding that the proposed watermarking system had the required effect in protecting an image. The five tests emulated the expected management attacks by a Cloud service provider and the artefact performance was sufficient that the watermark remained robust. This has implications for the redistribution of responsibilities for security management and for the type of technical system that can deliver rightful ownership protection.

## Discussion

Cloud computing introduces a range of risks a user has to reconcile with their appetite. The user also has expectations for privacy and ownership protection that may not be met in many Cloud computing environments. The present purchasing arrangements for services obscure the potential loss of control the user may experience. Sales agents are employed to sell the service and may not be informed of complex service arrangements. Service level agreements within and between service suppliers are service centric and have many interpretations across jurisdictional boundaries. As a consequence users have generally taken responsibility to provide security mechanisms such as encryption for their data. The approach has left a legacy of issues around the effectiveness of such measures and the viability of variation in a controlled environment. The research completed suggests that if service suppliers take responsibility for information security then the variation in security mechanism performance can be reduced and suitable mechanism may be tested by the service provider prior to use to assure user data control.

The research specifically focused on five management attacks that can be expected in a Cloud service environment. The artefact selected was a watermark that had been prepared with these attacks in mind. It had three layers and embedded security features to promote the longevity. The performance showed the torrid nature of policy driven attacks. No watermark escaped degradation and the best lost 30% of the intensity. This suggests that the problem identified is a serious issue and further work is required to assure robust preparation algorithms for future artefacts. The worst case lost almost 50% of the intensity suggesting that the nature of an image has an influence on performance. Further questions arise regarding the extent to which an artefact may be exposed to and in such an environment before the intensity drops below detection. Metrics such as duration, respective occurrences, pixel intensity and so on can be valuable indicators for forecasting an artefact robustness. In this research cloud technical attacks were out of scope and these can be investigated in future iterations of the research. The management attack results suggest that some images may lose further intensity when exposed to further attacks and reduce the positive impact of these findings. It can be anticipated that all of the managerial attacks will be present and some technical. In such an environment information regarding the artefact performance is required before a complete solution to the research problem is reported. However, the results give a strong indication that managerial attacks can be overcome and that the artefact has potential for further development. The suggested redistribution of responsibility for security to the service provider also places on them the responsibility to develop a robust solution that users may choose to use or used by default.

The research methodology has achieved the aim of answering the research question in a series of partial solutions and a forecasted further round of testing for technical attacks. As such the methodology has delivered against the six phases of activity. The applicability to IS security research has been demonstrated. The concept of Cloud security and relevant mechanism performance are still maturing in the literature. There are many gaps and big assumptions that have come from using security mechanisms from other environments in the Cloud. The Cloud represents a new context in which to design security solutions. In this paper we have taken one mechanism and a selected range of attacks to show how the DS framework can be applied for achieving IS security research. The DS framework has given the flexibility to try and to test assumptions and then when complete the ability to loop back and to seek improvement, answers to questions raised, and to address incomplete parts in this research. As such DS as a framework and a methodology is an effective approach for managing security mechanism research in new environments and contexts.

The issue of rightful ownership and inter jurisdictional issues surrounding the cloud will not go away. These are material concerns that have eroded trust in cloud services but may be negotiated by better understandings and mitigated by better application of security technologies to the new environment. We proposed a different system architecture to better fit the watermark security technology into the cloud environment and also built an artefact that has potential to fit the new environment. Such innovation may become common practice as cloud services move out of their infancy and greater trust is gained by more users. The users who unwittingly use cloud services by default also require assurance that their privacy and ownership is protected. Further research and development are required to grow the effective application of security technologies to the cloud environment.

# 5   Conclusion

The proposal has been to redistribute responsibility for watermarking to the service provider in order to achieve consistency in watermarking and effectiveness in security. The service provider may market the mechanism as a value added service or a default for users. In addition a watermarking process has been developed and tested to be robust in the cloud environment. The Design Science methodology has allowed an evaluation of the proposal and critical reflection on the research processes after moving the artefact through all six phases. The artefact can now be re-entered into phase three or phase two of the methodology for quality improvement and further evaluation. Within the limitations discussed the advocated solution returns an improved quality of control to the intellectual property owner.

## Copyright